\documentclass[12pt,preprint]{aastex}
\usepackage{natbib}
\usepackage{subfigure}
\usepackage{times}
\begin{document}

\title{Modeling the multiwavelength emission from G73.9+0.9: Gamma-rays from a SNR-MC interaction}

\author{Miguel Araya}
\affil{Centro de Investigaciones Espaciales (CINESPA) \& Escuela de F\'isica,\\ Universidad de Costa Rica}
\affil{San Jos\'e 2060, Costa Rica}
\email{miguel.araya@ucr.ac.cr}

\slugcomment{To appear in ApJ}

\begin{abstract}
G73.9+0.9 has been classified as a probable shell-type supernova remnant (SNR), although it has also been suggested that this object could be a pulsar wind nebula (PWN). Here, a broadband model of the non-thermal emission of G73.9+0.9 from radio to gamma-rays is presented. The model includes a new gamma-ray observation obtained with analysis of 7 years of data from the \emph{Fermi} LAT telescope. Above 200 MeV the source is detected with a significance of 13$\sigma$ and the spectrum of the radiation is best described by a power law with an index of $\sim$2.5. The leptonic mechanisms are hard to reconcile with the measured radio and gamma-ray SED. A PWN origin for the high-energy emission is also not very likely, due to the lack of detection of pulsars and of X-ray emission in the region, as well as from the shape of the gamma-ray spectrum. Given the possibility that the object is interacting with molecular clouds, a hadronic origin of the high-energy emission is more likely, and the spectral properties of the cosmic rays responsible for this radiation are derived.
\end{abstract}


\section{Introduction}
In the last decade, an increasing number of supernova remnants (SNRs) have been detected at gamma-ray energies. Establishing this type of objects as high-energy emitters is important as photon signatures could help solve the long-standing problem on the origin of Galactic cosmic rays, charged particles with energies up to $10^{15}$ eV. These particles are thought to experience acceleration in the shocks of SNRs which could transfer up to 10\% of the explosion kinetic energy ($\sim 10^{51}$ erg) to particles through the well-known mechanism of diffusive shock acceleration \cite[e.g.,][]{bel78,bla87}. Observations of synchrotron radiation from SNRs, from radio to X-rays \citep{got01,ber02,hwa02,rho02,lon03,vink03,ber04}, reveal the existence of high-energy electrons in these sites. 

Gamma-rays may be produced by high-energy particles mainly through the mechanisms of inverse Compton (IC) up-scattering of low energy ambient photons, non-thermal bremsstrahlung radiation and the decay of neutral $\pi-$mesons produced in inelastic collisions between accelerated ions and ambient nuclei. The hadronic mechanism may produce a gamma-ray spectrum which can be hard to distinguish from leptonic emission. Some SNRs are bright gamma-ray sources and are also known to interact with molecular clouds (MC) or produce high-energy particles that interact with these clouds \citep{abdo2010a,abdo2010b,abdo2010e,abdo2010f,abdo2009,ackermann2013,ajello2012,xing2014,xing2015,liu2015}. These observations support the idea that cosmic rays are produced in SNRs. The high-density material in MCs could provide abundant targets for gamma-ray production in a hadronic scenario, making SNR-MC interactions ideal targets to constrain the spectrum and energetics of cosmic rays. On the other hand, the origin of gamma-rays in younger remnants remains controversial. While the emission from Tycho's SNR or Cas A might be of hadronic origin \citep{abdo2010d,araya2010,araya2011}, others show gamma-ray spectra that is characteristic of IC emission \citep{ellison2010,yuan2011,tanaka2011,yuan2014}.

At high energies, SNRs have been observed by the \emph{Fermi} LAT, a pair conversion telescope with a field of view of 2.4 sr sensitive to gamma-rays from $\sim$ 20 MeV to more than 300 GeV \citep{atw09}. Here, a detailed analysis of LAT data in the region of the SNR G73.9+0.9 and broad-band modeling are presented. This object is located near the complex Cygnus region and it is seen in radio images as a compact nebulosity with a size of $\sim 27'$. It is classified as an SNR based on its polarization and nonthermal spectrum \citep{reich1986}. Its measured flux density is 9 Jy at 1 GHz and the photon spectral index is ($F_{\nu}\propto \nu^{-\alpha}$) $\alpha=0.23$ \citep{kothes2006}. The type of remnant is uncertain, the possibilities include a shell type remnant or a pulsar wind nebula. No pulsars were found in a search by Gorham et al. \citep{gorham1996} with a sensitivity of 0.3 mJy at 430 MHz and 1400 MHz, and another pulsar search by Lorimer et al. \citep{lorimer1998} also resulted in no detection.

The distance to the source is relatively unknown. Lozinskaya et al. \citep{lozinskaya1993} derive a distance of $1-2$ kpc based on optical observations. Based on the $\Sigma-D$ relation, a distance of $\sim 6$ kpc was determined by Leahy \citep{leahy1989}. Mavromatakis \citep{mavromatakis2003} conclude that a distance of 2 kpc is compatible with their optical observations of the source. Lorimer et al. \citep{lorimer1998} cite the study by Lozinskaya et al. \citep{lozinskaya1993} and use a distance of 1.3 kpc. Pineault et al. \citep{pineault1996} find an interesting HI region that shows a correlation with the remnant's morphology at a kinematic distance of 10 kpc, but this distance is probably less likely for G73.9+0.9. A distance closer to more recent estimates of 1.8 kpc is adopted here.

From their observations, Lozinskaya et al. \citep{lozinskaya1993} also estimate a shock expansion velocity of $200-300$ km/s and age of $11-12$ kyr for G73.9+0.9, for an ambient density of 10 cm$^{-3}$. They compare these properties to those of the well-known SNR IC 443 and find similarities.

A number of compact radio sources are seen near the remnant, most of which have spectra that are compatible with HII regions or dust pressumably heated by the shock \citep{pineault1990}. $^{12}$CO $J= 1 - 0$ observations show molecular clouds with spatial correlations with the radio emission from the SNR at small CO velocities (+1 to +3 km/s, Jeong et al. \citep{jeong2012}) to the north of the radio peak, although no direct evidence for an interaction has been reported. Other studies of the ambient near the SNR have concluded that there is an HI shell around it that are likely associated with it, and have also detected molecular cloud emission to the northwest of G73.9+0.9 \citep{sitnik2010}.

Mavromatakis \citep{mavromatakis2003} derive column densities that imply ISM densities of 0.5 cm$^{-3}$ at the distance of 2 kpc and conclude, based on the properties of the optical emission, that an extended radio blob seen to the north of the SNR is probably unrelated to it, but perhaps associated to a bipolar HII region. This had also been concluded by Pineault et al. \citep{pineault1996} after comparing the infrared properties of the SNR and the northern component, which are very different, as is the morphology of the HI gas in both regions. According to Mavromatakis \citep{mavromatakis2003}, the low ionization optical images further indicate an inhomogenous and patchy intestellar medium.

X-ray observations from the ROSAT All-Sky survey of the region show no especially prominent soft emission. Leahy \citep{leahy1989} analysed \emph{Einstein} observations and derived a 2$\sigma$ upper limit on the $0.2-4$ keV flux of $3.0\times10^{-13}$ erg cm$^{-2}$ s$^{-1}$. There is an archival XMM-Newton observation of the remnant that was proposed to search for a pulsar wind nebula but no obvious emission was found (private communication). At very-high energies (VHE), VERITAS set a 99\% upper limit for the source above 300 GeV \citep{theiling2009} of $2.33\times10^{-12}$ photons cm$^{-2}$ s$^{-1}$ with $\sim 600$ minutes of data (assuming a power-law spectrum, $dN/dE \propto E^{-\Gamma}$, with index $\Gamma=2.5$). The radio fluxes used in this work are taken from the literature \citep{reich1986,pineault1990,sun2011}

In this work, the gamma-ray observation is presented in Section \ref{data} and its properties explored. The spectral energy distribution (SED) is modelled in Section \ref{models}. Different possibilities for the origin of the emission are studied. Finally, the conclusions are given in Section \ref{conclusions}.

\section{LAT Data}\label{data}
LAT data gathered from the beginning of the mission, 2008 August, to 2015 July were analysed with the most recent software, {\small SCIENCETOOLS} version v10r0p5\footnote{See http://fermi.gsfc.nasa.gov/ssc}, released 2015, June 24, with the latest reprocessed ``Pass 8'' photon and spacecraft data and the instrument response functions P8R2\_SOURCE\_V6P. The Pass 8 is an effort to improve the reconstruction algorithms applied to the data and includes an overall improvement in the understanding of the detector \citep{atw13}.

Photons with energies from 200 MeV to 100 GeV are selected within a 14$^{\circ}\times$14$^{\circ}$ square region of interest (ROI) centered at the position of G73.9+0.9, (J2000) $\alpha$ = 303.54, $\delta$ = 36.20. Below 200 MeV the steeply falling effective area and broadening of the energy dispersion and PSF present significant challenges in the analysis, and above 100 GeV the drop in the number of photons becomes an issue.

The data selection is limited to zenith angles less than 90$^{\circ}$ to remove photons from the Earth limb and the event class is cut to keep only the SOURCE class events as recommended. Time intervals are selected when the LAT was in science operations mode and the data quality was good. The data is binned in 30 logarithmically spaced bins in energy and a spatial binning of $0.^{\circ}1$ per pixel is used.

The analysis includes all sources listed in the latest \emph{Fermi} LAT catalog (hereafter 3FGL, Acero et al. \citep{acero2015}), as well as the standard diffuse Galactic emission model and the isotropic model that accounts for the extragalactic gamma-ray background and misclassified cosmic rays, both of which are provided with the analysis tools. The normalizations of these spatial templates are left free during the likelihood fit, as well as the normalizations of the cataloged sources located less than 6$^{\circ}$ from the center of the ROI, to account for the broad PSF at low energies. The spectral parameters of all the sources are kept fixed during the fit.

The properties of LAT sources are obtained by a maximum likelihood analysis \footnote{See http://fermi.gsfc.nasa.gov/ssc/data/analysis/scitools/binned\_likelihood\_tutorial.html}, which estimates the probability of obtaining the data given a source model and fits the source parameters to maximize this probability. The unidentified source candidate 3FGL J2014.4+3606, located about $0^{\circ}.1$ from the center of the ROI, is reported in the 3FGL catalog with a significance of 4.5$\sigma$. This source candidate is not included in the model in order to accurately study the morphology of the residuals and the significances of new sources seen with the larger data set used here.

The bright LAT pulsar 3FGL J2021.1+3651 is found $1.^{\circ}6$ from the center of the ROI. This source shows a spectrum with a cut off at 3 GeV. Other nearby sources include the FSRQ 3FGL J2015.6+3709 ($1.^{\circ}1$ from the ROI center, also known as VER J2016+371) and 3FGL J2017.9+3627 (located $\sim 1^{\circ}$ from the SNR), believed to be associated to the VHE source MGRO J2019+37. These sources have LAT spectra that are described by a logParabola. In order to investigate the effect on the residual emission near these sources, fits are done in different energy intervals keeping these source parameters free, and the resulting residuals maps obtained after subtracting the best-fit model to the data are compared to those obtained when their spectral parameters are fixed to the values reported in 3FGL. The results are very similar in both cases. The spectral analysis in Section \ref{spectrum} was also repeated keeping the spectral shape of nearby sources free without significant differences in the results. The results of the analysis reported here was then obtained with spectral parameters of all sources fixed to the values found in the LAT catalog.

In order to remove the emission of nearby gamma-ray sources as well as the diffuse backgrounds, the tool gttsmap is used above 3 GeV in a 1.2$^{\circ}\times$1.2$^{\circ}$ region around the center of the ROI. For each pixel of the map the test statistic (TS) is evaluated. The TS is defined as $-2$ log($L_0/L$), where $L_0$ and $L$ are the maximum likelihood values for the null hypothesis and for a model including an additional point source at the location of that pixel, respectively \citep{mat96}. The detection significance of a source is given by $\sqrt{\mbox{TS}}$ \cite[however see][for some caveats]{protassov2002} and thus a TS value of 25 is considered the threshold for source detection.

\begin{figure}[ht]
\centering
\includegraphics[width=14cm,height=9cm]{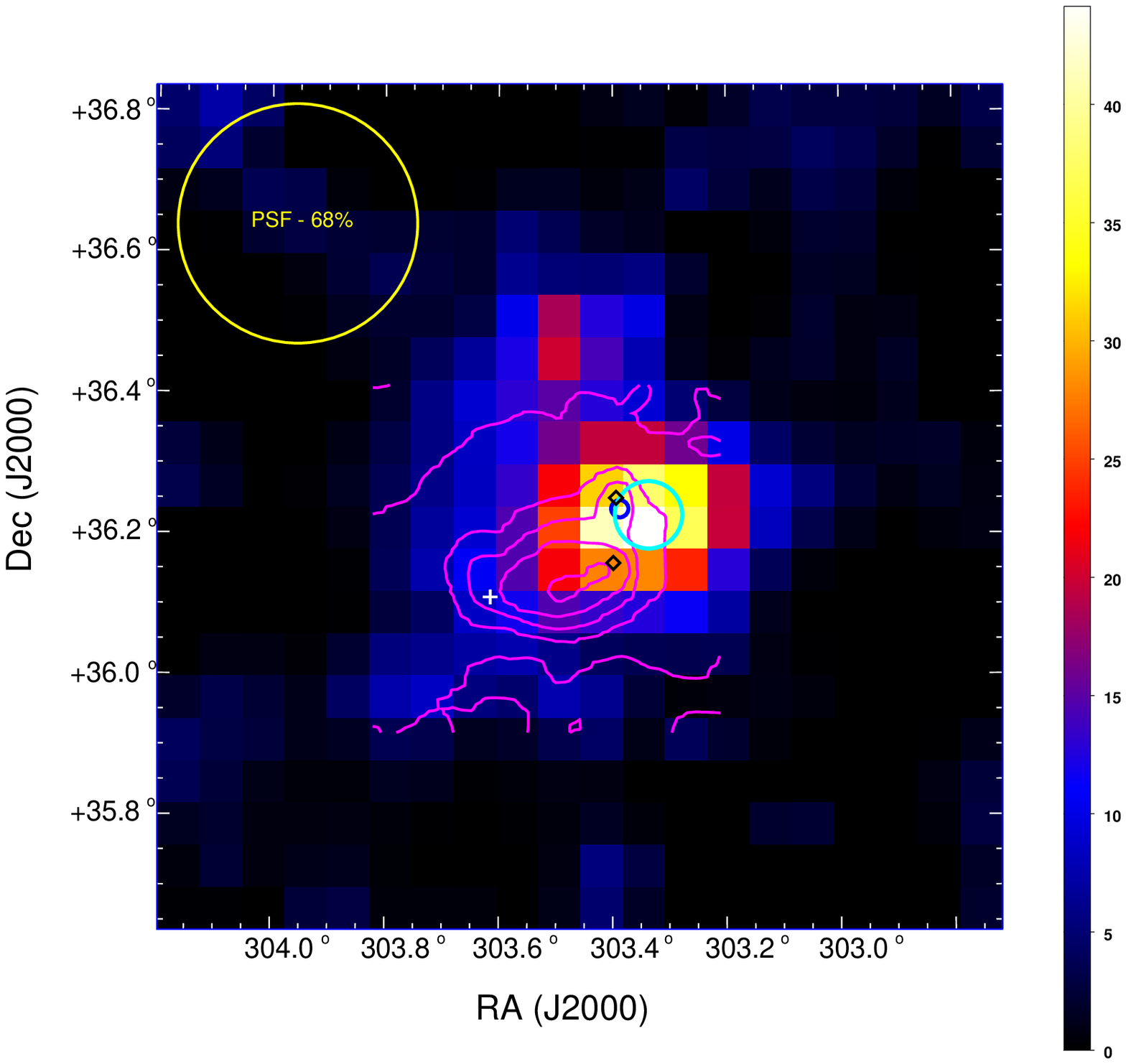}
\caption{Significance map of the gamma-ray emission around G73.9+0.9 output from gttsmap above 3 GeV, in units of test statistic (TS, see text). The map has a dimension of 1.2$^{\circ}\times$1.2$^{\circ}$ and a resolution of 0$^{\circ}$.06 per pixel. The magenta contours mark the non-thermal radio emission of the SNR for five equally-spaced levels between 0 and 0.3 Jy/beam, taken from the 4850 MHz Green Bank survey \citep{condon1994}. The center of the cyan circle indicates the best-fit position (and its radius the positional uncertainty) of the gamma-ray source, the white cross indicates the position of the previous 3FGL source candidate and black diamonds denote the positions of radio knots which are likely HII regions. The blue circle marks the location of the peak of emission from a molecular cloud, obtained from $^{12}$CO $J=1-2$ observations by Jeong et al. \citep{jeong2012}. No other 3FGL source is found within this map and no known blazars are seen in it either \citep{mass2009}. The yellow circle shows the approximate size of the LAT PSF for 68\% containment of FRONT+BACK events.\label{fig1}}
\end{figure}

Fig. \ref{fig1} shows the resulting significance map around G73.9+0.9 as well as the locations of other objects in the region. The PSF of the LAT suggests that the excess found here is consistent with a point source, as it would be expected for an object with the extension of SNR G73.9+0.9, or any structure within it. In order to find the position of the new source the tool gtfindsrc is applied to the same data set above 3 GeV, the resulting best-fit coordinates are (J2000) $\alpha$ = 303.337, $\delta$ = 36.226 with a 68\% confidence level positional error radius of 0$^{\circ}$.05. As an additional test, residuals maps are also obtained at lower energies and the excess observed seems consistent with a point source at this position, indicating that there are no further sources in the region having a softer spectrum that would not have shown up in the TS map above 3 GeV.

The map shows a separation of $\sim 0^{\circ}.25$ between the new best-fit position and the position of the source candidate reported in the 3FGL catalog, obtained with a smaller data set but also with different analysis software versions and diffuse background models. The significance of the new source reported here is also different ($\sim 6\sigma$ above 3 GeV, but much higher for the whole energy range, see below). A satisfactory substraction of the emission is also seen in the residuals map once this source is added to the model.

\subsection{Spectrum of the source}\label{spectrum}
Once the new source position and morphology are known, its spectrum is studied above 200 MeV with the likelihood technique. Different spectral shapes are probed in order to compare the resulting TS values: simple power-law, broken power-law, logParabola, and power-laws with exponential and `super' exponential cut offs. It is found that models that require more free parameters with respect to the simple power-law result in similar TS values, which are (in the same order): 177, 184, 180, 182 and 181. Thus, the fits with all these functions correspond to an improvement at the $\sim 2\sigma$ level with respect to the simple power-law, which is not very significant. The best-fit spectral shape adopted in the rest of the analysis is a simple power-law. The values of the integrated source flux for these different spectral shape fits are very similar.

Above 200 MeV the spectral index of the source is $(2.51 \pm 0.07_{stat})$ with an integrated flux above this energy of $(1.87\pm0.21_{stat})\times 10^{-8}$ cm$^{-2}$s$^{-1}$, and the overall significance of the source is quite high, $13\sigma$. The corresponding luminosity above 1 GeV is $(2.7\pm 0.3_{stat})\times 10^{33}\,d_{1.8}^2$ erg s$^{-1}$, where $d_{1.8}$ is the distance in units of 1.8 kpc. This value is very similar to the luminosity of other dim gamma-ray remnants such as the Cygnus Loop, G296.5+10.0 and HB9 \citep{katagiri2011,araya2013,araya2014}. The source TS drops to 3 for events above $\sim 6$ GeV.

In order to construct a gamma-ray SED the data is binned in 9 logarithmically-spaced intervals from 0.2 MeV to 100 GeV, and a likelihood fit is applied within each keeping the spectral index of the source fixed to the global value. Whenever the TS of the source falls below 9 a 95\% confidence level upper limit is derived in the corresponding interval. Several systematic uncertainties are taken into account in the SED fluxes for each bin. According to the analysis of Pass 8 LAT data, the uncertainty on the effective area is about 5\% and the systematic uncertainty on the PSF is always less than 5\% in the energy range used here. These uncertainties are added in quadrature to the flux statistical errors in the SED.

\subsection{Variability analysis}\label{timing}
\begin{figure}[ht]
\centering
\includegraphics[width=14cm,height=8cm]{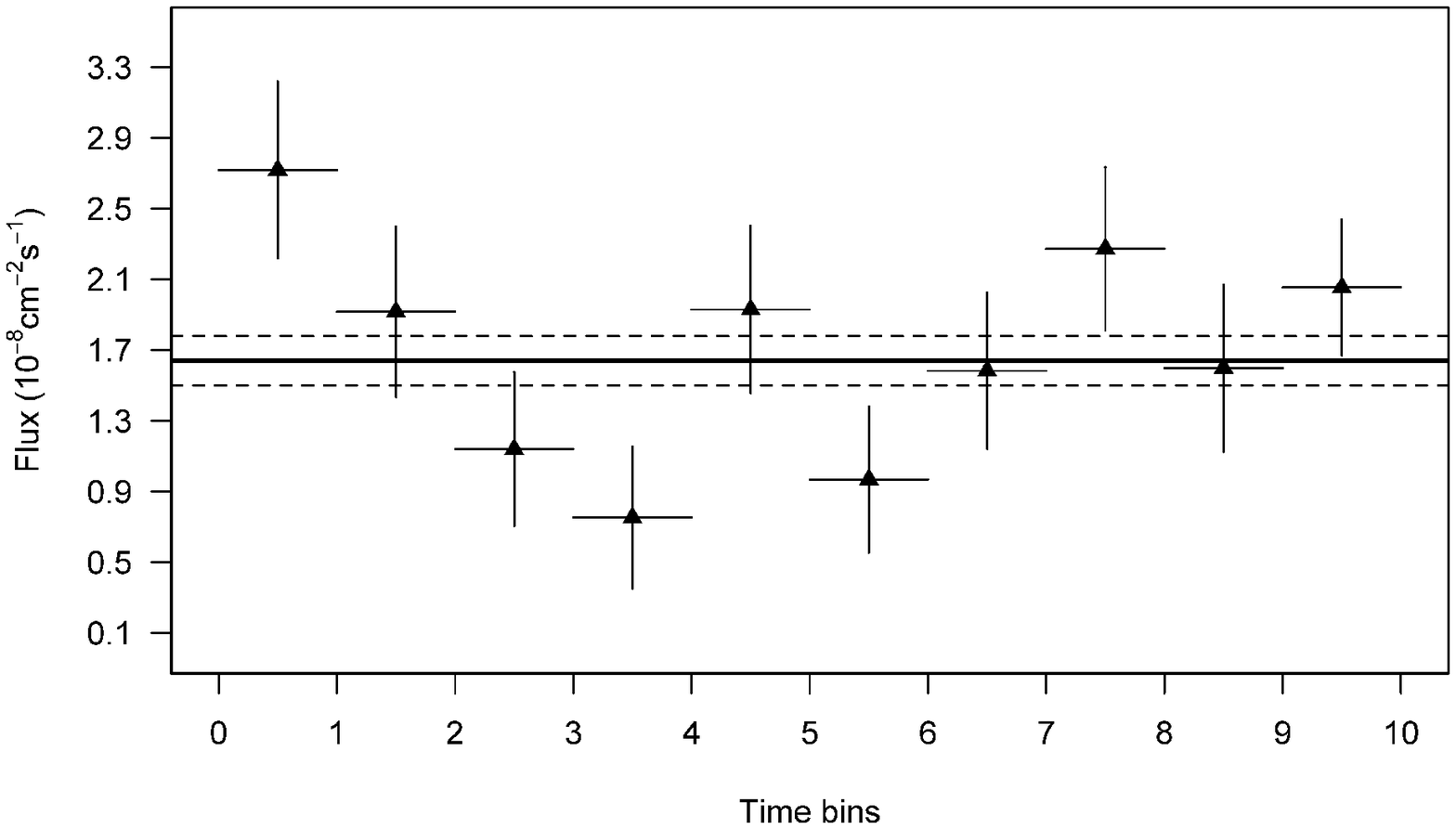}
\caption{Lightcurve of the source seen at the location of G73.9+0.9. The fluxes are calculated for 10 time bins in the energy range $0.2-100$ GeV and show only statistical errors. The solid line corresponds to the average flux and the dashed lines mark the $1\sigma$ error interval around this value.\label{fig2}}
\end{figure}

Extragalactic gamma-ray sources are known to be highly variable. In order to search for variability, the data is binned in ten $\sim8$-month bins and the likelihood analysis is applied to estimate the flux of the source seen in the direction of G73.9+0.9. The normalizations of the other sources in the model are kept fixed to the values found for the fit to the 7-yr data set, as well as the spectral index of G73.9+0.9. Fig. \ref{fig2} shows the resulting fluxes for each bin. A fit to a constant value gives an average flux of $(1.64\pm 0.14)\times10^{-8}$ cm$^{-2}$s$^{-1}$ ($1\sigma$ error), and a reduced $\chi^2$ of 1.9.

In order to quantify the probability of changes in the flux with time, a variability index is calculated \citep{abdo2010c} and the result is 17.2 considering statistical errors on the data only. Varibility is considered probable when this index exceeds the threshold of 21.7, corresponding to 99\% confidence in a $\chi^2$ distribution with 9 degrees of freedom. Adding systematic errors would result in a lower index. It can be said that the source shows no significant long term variability.

\section{Discussion}\label{models}
\emph{Fermi} LAT data confirm the existence of significant gamma-ray emission from an unresolved source at the location of G73.9+0.9, an object whose nature is still debated. The spectrum and properties of the source are now studied in the context of multi-wavlength obervations considering several scenarios. 
\subsection{A pulsar or pulsar wind nebula}
Pulsars are an important class of gamma-ray sources. No radio pulsars have been found in the region of the SNR. Most pulsars have gamma-ray spectra that can be fit with a flat power-law that cuts off exponentially at energies of a few GeV \citep{abdo2013,ackermann2012,lee2012}, while the spectrum of the gamma-ray source detected here has been shown to be described by a simple power-law. Some pulsars receive a ``kick'' during the supernova explosion. The angular separation from the position of the gamma-ray source and the peak of the radio shell is $\sim0.^{\circ}14$. Even if the remnant is assumed to be relatively young (age $\sim1$ kyr), which is unlikely given the lack of detection of X-ray emission, the corresponding pulsar transverse velocity is 4500 km/s for a distance of 1.8 kpc, which is high but still possible; however, no pulsars have been found in the region.

Pulsars transfer part of their rotational energy to a relativistic magnetized wind (mostly electron-positron pairs). When this wind slows down abruptly it creates a termination shock where particle acceleration occurs, resulting in a pulsar wind nebula (PWN). PWNe are seen from radio to gamma-rays. Their non-thermal emission is dominated by synchrotron and inverse Compton from high-energy leptons, although hadronic processes may also contribute \citep{cheng1990,amato2003}.

G73.9+0.9 has several radio properties that are similar to those of PWNe such as a low spectral index. However, PWNe detected by the LAT typically have young, bright pulsars, and are also seen at TeV energies \citep{abdo2011}. Hard X-ray and gamma-ray observations of PWNe have the potential to constrain, among other things, the spectral break and maximum energy of synchrotron-emitting particles. The radio spectrum of a PWN is hard ($F_{\nu}\propto \nu^{-\alpha}$, $\alpha=0-0.3$) and softens at higher energies, and in the X-ray band $\alpha > 1$. The synchrotron spectrum of the Crab Nebula, for example, may be described as power-laws with breaks around $10^{13}$ Hz, $10^{15}$ Hz and 100 keV and only the peak at VHE is attributed to IC emission. PWNe usually have relatively high radio polarization fractions (from 30\% to 50\% at 1 GHz), while for G73.9+0.9 it is closer to the value found for SNRs, around 1\% to 5\% at $\sim5$ GHz \citep{sun2011}.

\begin{figure}[ht]
\centering
\includegraphics[width=14cm,height=7.5cm]{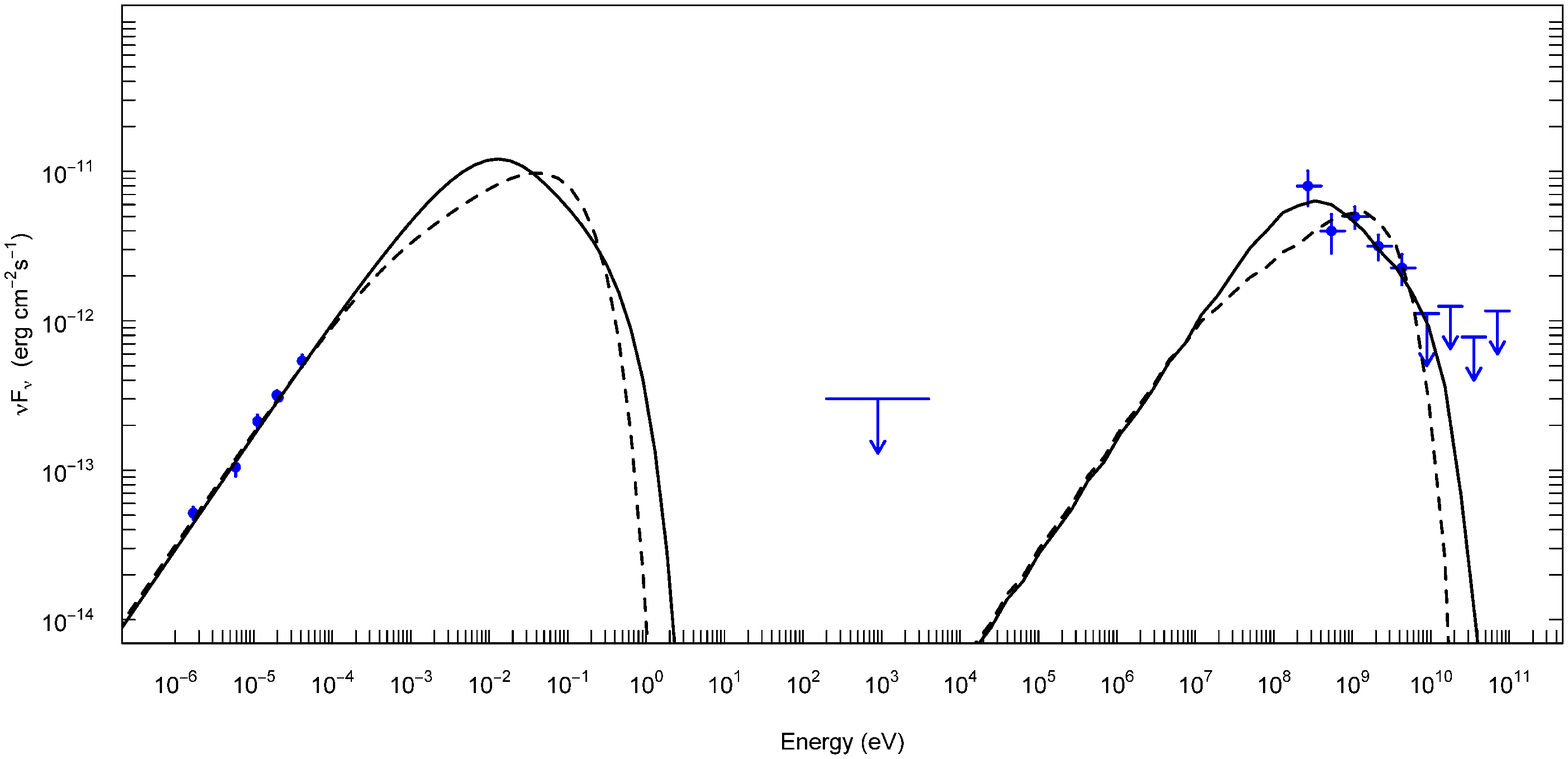}
\caption{SED of G73.9+0.9 and two simple one-zone leptonic models in the PWN scenario (synchrotron plus IC-CMB). The gamma-ray points and upper limits are from the LAT observation in this work. The particle distribution in both models is a broken power-law with indices and break particle energies of: 1.46, 4.0, 204 GeV (solid lines); and 1.46, 2.2, 50 GeV (dashed lines).\label{fig3}}
\end{figure}

PWNe can also have a wide range of nebular magnetic fields \cite[$\sim5$ $\mu$G to $>1$ mG, e.g.,][]{reynolds2012} and show a variety of break frequencies in their radio spectra. Fig. \ref{fig3} shows the SED from radio to gamma-rays with two possible models that include synchrotron emission in a magnetic field of 4 $\mu$G and IC scattering of CMB photons (this low field value is only a lower limit, as the IC emission from electrons up-scattering other ambient photons is not considered). The radio fluxes correspond to integrated emission of G73.9+0.9 \citep{reich1986,pineault1990,sun2011}. The \emph{Einstein} UL at X-ray energies implies that the GeV emission cannot be produced by synchrotron emission from the same radio-emitting particles. The models are shown for broken power-law particle distributions with different indices and break and maximum electron energies: 1.46, 4.0, 204 GeV and 1 TeV, and 1.46, 2.2, 50 GeV and 600 GeV, respectively. A lepton energy density of $\sim10^{-10}$ erg cm$^{-3}$ is used in both cases, which is much higher than the resulting magnetic field energy density ($6.4\times10^{-13}$ erg cm$^{-3}$). The total particle energy content is $\sim4\times10^{48}$ erg and the models assume a source distance of 1.8 kpc. Contributions from synchrotron self-Compton emission are negligible.

No known LAT PWNe show a peak of the IC emission in the LAT energy range. In a search for LAT PWNe, the confirmed PWNe and the candidates were always associated with young and powerful pulsars with spin-down power between $10^{36}$ and $10^{39}$ erg s$^{-1}$ \citep{acero2013}, although the search was biased as the objects studied were previously selected for having associated VHE emission. In a systematic exploration of models of Crab-like PWNe, Torres et al. \citep{torres2013} studied a variety of parameters affecting the SED of these objects, including the age. From their work, no case is seen where a source would show a very low X-ray flux ($< 10^{-13}$ erg cm$^{-2}$ s$^{-1}$) and a negative SED slope at GeV energies as found here for G73.9+0.9, at least up to a source age of 9 kyr, similar to the possible age of this object.
\subsection{An AGN?}
Active Galactive Nuclei (AGN), and blazars in particular, are an important class of gamma-ray sources. 3FGL includes 1745 sources associated with AGN (58\% of all 3FGL sources), and 98\% of them are blazars or candidate blazars, the rest being radio galaxies, Seyfert galaxies and other types of AGN. Most of the gamma-ray emitting AGN are bright sources of radio emission. No blazars are found in the region near G73.9+0.9 \citep{mass2009} and no other AGN types are found in the Million Quasars Catalog \citep{flesch2015} (version 4.5, May 2015) near the gamma-ray source. The high-energy emission is also found to be consistent with steady emission in $\sim$yearly time scales (Fig. \ref{fig2}).

\subsection{SNR-MC interaction}
There is no direct evidence of an interaction of G73.9+0.9 with a MC and more studies are necessary to confirm or reject this scenario, but the morphological properties of the ambient clouds, as observed by Jeong et al. \citep{jeong2012}, point to this possibility. In particular, there is an interesting compact MC within the best-fit position of the gamma-ray emission, as can be seen in Fig. \ref{fig1}.

\begin{figure}[ht]
\centering
\includegraphics[width=14cm,height=7.5cm]{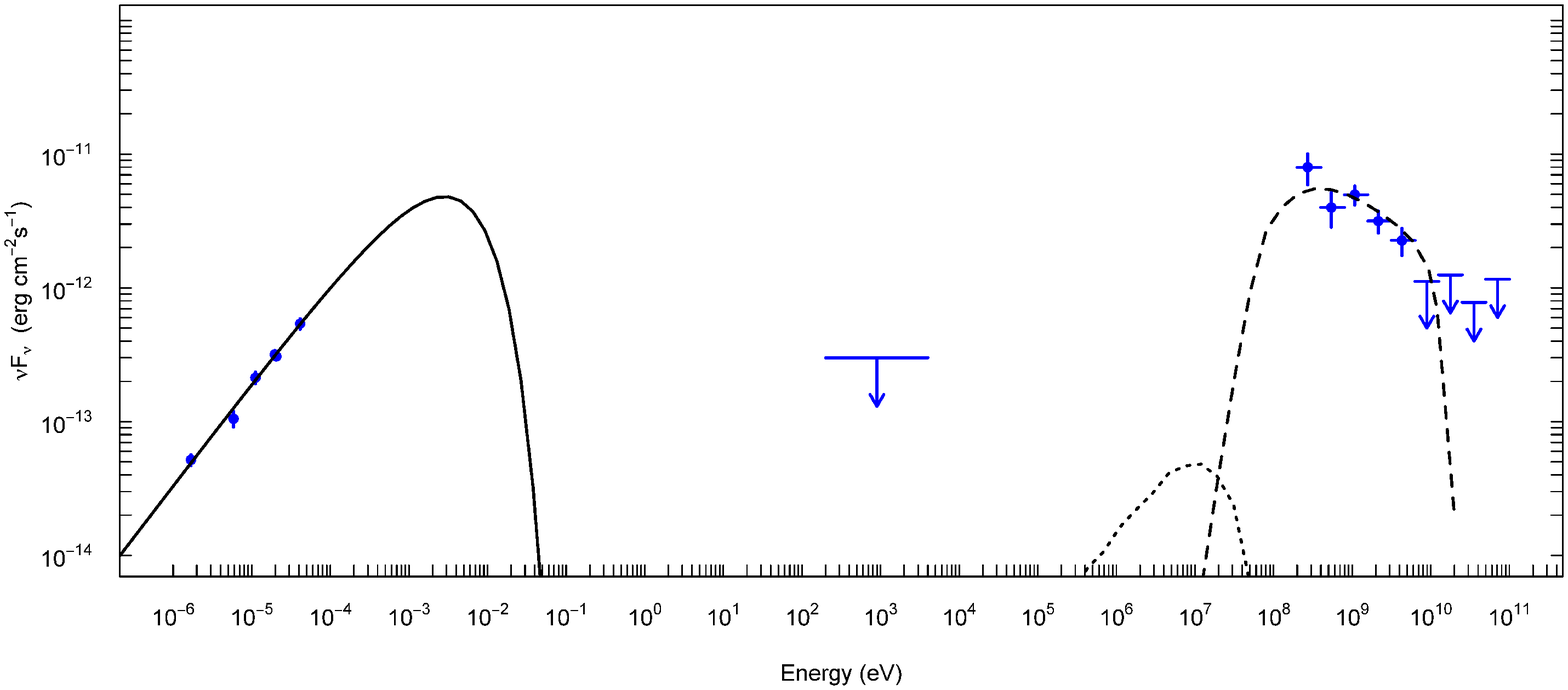}
\caption{Hadronic model for the SED of G73.9+0.9. The solid line corresponds to synchrotron emission (magnetic field of 30 $\mu$G, maximum electron energy 50 GeV), the dashed line is the gamma-ray flux from cosmic-ray interactions (ambient density $n_p=1$ cm$^{-3}$, total cosmic ray energy 10$^{50}$ erg) and the dotted line IC scattering of CMB photons by the radio-emitting electrons.\label{fig4}}
\end{figure}

Fig. \ref{fig4} shows an SED model for which the dominant contribution to the gamma-ray emission is hadronic, calculated with a recent parametrization of the cross section \citep{kafexhiu2014} for these interactions. The cosmic ray distribution used to calculate the emission is a simple power-law in momentum of the form $p^{-2.5}$. The use of this proton distribution is motivated by the shape of the gamma-ray spectrum. Hadronic gamma-ray emission is expected to increase rapidly around photon energies of 100 MeV and then follow the particle distribution above $\sim 1$ GeV. The model is calculated for a total proton energy of $10^{50}\left(\frac{n_p}{1\,\mbox{cm$^{-3}$}}\right)^{-1}$ erg, where $n_p$ is the target proton density. Since MCs can have a wide range of densities (including $n_p>>1$ cm$^{-3}$), the required total particle energy is realistic. The proton power-law spectrum in the model cannot extend beyond a particle energy of $15-20$ GeV, based on the LAT upper limits.

Even though the SED obtained with this model reproduces the observed SED, the choice of a broken power-law for the particle distribution can also succesfully explain the data. Gamma-ray spectra observed in systems with SNR-MC interactions are sometimes modeled with a broken power-law \cite[e.g.,][]{abdo2009,abdo2010e,abdo2010f,ohira2011,tang2011}. In this scenario, a proton distribution with a break momentum of $\sim$3 GeV c$^{-1}$ and indices of 1.5 below (as found for the leptonic component) and 3 above this break, with a total cosmic ray energy of $0.8\times10^{50}\left(\frac{n_p}{1\,\mbox{cm$^{-3}$}}\right)^{-1}$ erg, can reproduce the data as well. These parameters, however, are not very well constrained, as also lower break momenta and lower indices above the break (i.e., harder proton distributions) are consistent with the data. According to previous work, the break momentum depends on the magnetic field, ion density and the frequency of ion-neutral collisions \citep{malkov2011}, but these parameters are unknown.

For the model shown in Fig. \ref{fig4}, the leptonic component is obtained with a simple power-law for the electron distribution (index 1.46 and maximum electron energy of 50 GeV) with a total energy in leptons of $1.5\times10^{47}$ erg and a magnetic field of 30 $\mu$G. If the field is closer to the average Galactic value, $\sim$3 $\mu$G, the corresponding total lepton energy would be $2.8\times10^{48}$ erg. Such low fields would still be sufficiently capable of confining protons with energies of tens of GeV producing the gamma-rays.

\subsection{Other leptonic mechanisms: bremsstrahlung and IC emission}
\begin{figure}[ht]
\centering
\includegraphics[width=14cm,height=7.5cm]{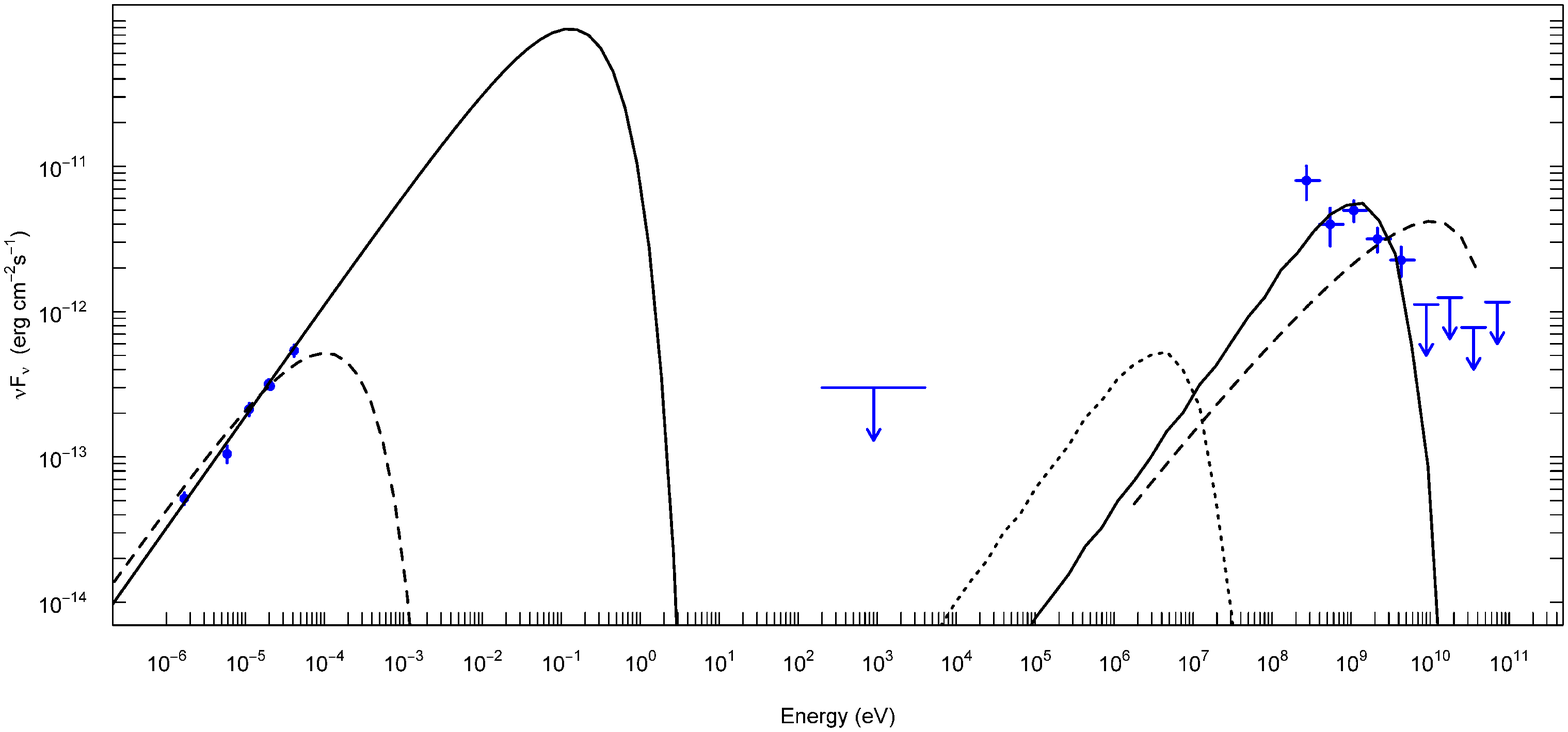}
\caption{Leptonic models for the SED of G73.9+0.9. The solid lines correspond to synchrotron and IC-CMB in an IC-dominated scenario (a magnetic field of 12 $\mu$G and maximum electron energy of 500 GeV are used). The dashed lines are synchrotron emission (maximum electron energy 30 GeV, magnetic field 3 $\mu$G) and non-thermal bremsstrahlung (target ambient density $n_p=1$ cm$^{-3}$) in the scenario where this latter emission dominates the gamma-ray flux (and for which the dotted line shows the corresponding IC-CMB flux level).\label{fig5}}
\end{figure}

It is difficult to explain the LAT fluxes below 400 MeV with IC-CMB emission from electrons following a power-law distribution with index 1.46. The magnetic field used is 12 $\mu$G and the maximum electron energy is 500 GeV. The synchrotron SED emission would peak around infrared frequencies ($\sim 3\times10^{13}$ Hz) at the level of $\sim 9\times 10^{-11}$ erg cm$^{-2}$ s$^{-1}$. This scenario can be seen in Fig. \ref{fig5}.

The relatively hard electron spectrum in this source makes a bremsstrahlung-dominated scenario for the gamma-rays very unlikely. Fig. \ref{fig5} also shows this case for an ambient target density $n_p=1$ cm$^{-3}$. The predicted gamma-ray flux could become somewhat similar to the level measured only when the maximum electron energy is well below $\sim 30$ GeV, a value for which the resulting radio spectrum starts to deviate from the data. The bremsstrahlung scenario can be safely discarded for the gamma-rays in G73.9+0.9, as it is unable to explain the gamma-ray data.

\section{Conclusions}\label{conclusions}
Gamma-ray emission has been found in the region of G73.9+0.9 with cumulative observations by the LAT onboard \emph{Fermi}. The spectrum of the emission can be best described with a power-law with index 2.5 in the interval where the signal is significant, between 200 MeV and $\sim$6 GeV. No other known possible gamma-ray emitter is seen at or near the location of the point-like source. Being in the region where there is a possible SNR, it is natural to explain the origin of the gamma-rays as due to interactions of high-energy particles with the ambient gas or radiation fields, but there are other possibilities.

Several scenarios were considered to explain the origin of the emission. There are no known pulsars in the region. The gamma-ray spectrum of pulsars usually shows a cutoff or curvature around a few GeV, which is not seen for this source. Also, there are no AGNs that are known in the direction of G73.9+0.9, and the flux from the source was shown to be consistent with steady emission in a time scale of years. The lack of detection of X-ray emission and the gamma-ray spectral shape are in conflict with the PWN scenario for the broadband SED, according to models for PWN with ages up to 9 kyr.

Due to all these constraints, it is more likely that the gamma-rays originate from interactions of high-energy particles accelerated in the SNR. The shape of the SED implies that the IC scenario is not very probable. On the other hand, the bremsstrahlung scenario is inconsistent with observations and can be discarded.

From the spatial correlation between the SNR and CO emission from a molecular cloud, whose presence would be ideal to enhance the gamma-ray luminosity from hadronic interactions, this possibility seems more favorable. CO emission is indeed seen within the best-fit position of the gamma-ray source. The GeV SED of this source is also consistent with a hadronic scenario and, within the uncertainties, the particle population responsible for the emission can be described by either a simple power-law (index of 2.5) or a broken power-law in momentum. Broken power-law distributions have been observed for the particles when the shock of a SNR interacts with a MC \cite[e.g.,][]{abdo2009,abdo2010e,abdo2010f}. The steepening in the distribution might be due to poor particle confinement in regions with low ionization \cite[e.g.,][]{ptuskin2003,malkov2011}. If, as is the case for the leptonic population, a hadronic index of $\sim$1.5 is used for the first component of the power-laws, the required index is 3 above a particle momentum break of 3 GeV/c, however, these parameters are not well constrained by the data.

\acknowledgments
This research has made use of NASA's Astrophysical Data System, the SIMBAD data base and the MILLIQUAS Catalog, Version 4.5. Financial support from Universidad de Costa Rica, through Vicerrector\'ia de Investigaci\'on and Escuela de F\'isica, is acknowledged. The author thanks I. Jeong for providing the CO data for the region of G73.9+0.9.

\end{document}